\documentstyle[11pt,aasms4,rotating]{article}

\def\simlt{\lower.5ex\hbox{$\; \buildrel < \over \sim \;$}}
\def\simgt{\lower.5ex\hbox{$\; \buildrel > \over \sim \;$}}

\def\d{{\rm d}}

\def\gcm3{{\rm\,g\,cm^{-3}}}
\def\ncm3{{\rm\,cm^{-3}}}

\def\>{$>$}
\def\<{$<$}

\lefthead{Melia et al.}
\righthead{}

\begin{document}
\vskip 0.5in
\title{\bf Energy Flows in the Jupiter-Io System}

\author{\altaffilmark{1} Siming Liu$^{1}$}

\affil{$^1$Key Laboratory of Dark Matter and Space Astronomy, Purple Mountain Observatory,
CAS, Nanjing, China, 210008}




\begin{abstract}

With the laws of mass conservation, momentum conservation and energy conservation, incorporating
the processes of neutral gas ionization and ion diffusion, we develop a self-consistent model for
the bright ribbon --- the most prominent feature in Io's plasma torus. The model parameters are
well constrained by earlier {\it in situ} observations with the Galileo and Voyager spacescrafts.
Our model calculation indicates that the total power dissipated inside the torus is 3.6 times
bigger than the total power transported to Jovian ionosphere via Birkeland current. The power
dissipation inside the torus is relatively uniform. Most of the power transportation associated
with the Birkeland current, however, is localized near the flux tube of Io. With a
height-intergrated conductivity of 0.15$\,$mho in Jovian ionosphere, consistent with earlier
aeronomy models, the model gives a reasonable fit to the recent observations of the FUV Io tail on
Jupiter. Extra mass loading near Io is required in the model. This excess of mass injection into
the plasma torus may explain the Io-correlated energy source revealed by earlier EUV observations
with the Voyager spacecrafts. Further investigation of the model implications is warranted.

\end{abstract}




%

\section{Introduction}

The coupling between magnetosphere and ionosphere in solar planets reveals many important
physical processes, which may have significant implications on distant astrophysical systems
[Mauk, Anderson and Thorne, 2002]. Since first detected in 1964 [Bigg 1964], the Io correlated
Jovian decametric bursts have been well studied [Goldreich and Lynden-Bell 1969; Riihimaa 1968;
Imai, Wang and Carr 1997]. The implied interaction between Jupiter and Io has been attracting
attention in the community for more than two decades [Dessler 1983; Belcher 1987].

Pioneer 11 first detected a plasma concentration near Io [Fillius, McIlwain and Mogro-Campero
1975], which was later revealed by Voyager 1 as a plasma torus [Broadfoot et al. 1979]. The
powerful EUV emission from this Io's plasma torus ($\sim 3\times 10^{12}$ W) has posed a
theoretical challenge since it was confirmed by Voyager 2 [Sandel et al. 1979]. Several models
have been developed to account for this, among which are the unipolar inductor model [Goldreich
and Lynden-Bell 1969, Belcher 1987], the Alfv\'{e}n wing theory [Drell, Foley and Ruderman 1965;
Neubauer 1980] and the model with an extended neutral cloud distributed in the torus [Brown and Ip
1981; Trauger 1984], with the last one being favored both theoretically [Barbosa, Coroniti and
Eviatar 1983; Barbosa 1994; Shemansky 1980] and observationally [Smyth and Shemansky 1983; Skinner
and Durrance 1986]. Although all the models achieve in explaining certain observations, a global
picture of the Jupiter-Io coupling system, which unifies these models, does not exist.

Significant progresses were made in recently years with the detection of Io's footprint on Jovian
ionosphere both at IR with the NASA IRTF [Connerney et al. 1993; Satoh and Connerney 1999] and at
UV with the Hubble Space Telescope (HST) [Clarke et al. 1996]. This is especially the case for the
newly reported downstream emission of the Io's footprint [Clarke et al. 2002] since it indicates
strong coupling between Jupiter's ionosphere and Io's plasma torus over an extend region [Hill and
Vasyli\'{u}nas 2002;  G\'{e}rard et al. 2002]. Now observations exist on both ends of this
coupling system. Combining the recent HST observations on the Io's footprint with the Galileo
spacecraft's {\it in situ} measurements [Frank et al. 1996; Kivelson et al. 1996; Gurnett et al.
1996], we develop a self-consistent model for one of the key features in the Io's plasma
torus---the bright ribbon [Trauger 1984; Dessler and Sandel 1993; Schneider and Trauger 1995;
Volwerk et al. 1997].

Our basic model assumption is that, led by Io, the extended neutral cloud is concentrated within a
narrow region of a few Io's radii. Then a self-consistent hydrodynamical model is developed for
such a cloud with ionization and plasma diffusion processes parameterized. The ionization of this
neutral gas will not only energize the plasma around it, which eventually lead to the formation of
the bright ribbon feature detected in the torus, but also induce significant Birkeland
currents which close in the northern and southern Jovian ionospheres, producing the Io's footprint
and its downstream emission. Thus correlation is expected between emissions from the ribbon and
that from its footprint. In the paper, we mainly study the energy flows in the system. Our model
calculation indicates that the energy transportation to Jovian ionosphere via Birkeland current,
which is localized near Io's flux tube, is comparable to the energy dissipation in the plasma
torus, which is relatively uniform along the ribbon. Future coordinated observations will be able
to test the model.

In \S\ \ref{equas}, we briefly discuss the basic ideas for introducing such a model and develop
full dynamical equations for the ribbon. The observational constraints on the model parameters are
discussed in \S\ \ref{para}. We then show that there is a very simple analytical solution to the
dynamical equations. In \S\ \ref{fit}, we calculate the model predicted energy flows and compare
them with observations. Our conclusions are drawn in \S\ \ref{con}.

\section{Model Description and Basic Equations}
\label{equas}

There is much observational evidence for the existence of an extended neutral cloud in the Io's
plasma torus [see e.g. Smyth and Shemansky 1983; Skinner and Durrance 1986]. Theoretically,
Shemansky [1980] first pointed out that the mass loading around Io is so insufficient that the
heating of the plasma induced by it cannot balance the cooling due to EUV emission from the plasma
torus. This conclusion was later confirmed by the Galileo {\it in situ} measurement [Bagenal
1997]. Russell and Huddleston [2000] recently showed that a small mass loading rate near Io is
also consistent with observations of the Jupiter-Io system made over the past three decades. This
small mass loading rate near Io then suggests an extended mass loading region, which may
associated with the extended neutral cloud observed.

A narrow ribbon in the Io's plasma torus was first reported by Trauger [1984] with Earth-based
optical observations. Dessler and Sandel [1993] later showed that the strong EUV emission from the
torus detected by the Voyagers is confined within a region less than $9\,R_I$ in width, where
$R_I= 1.8\times 10^6\,$m is the radius of Io, consistent with Galileo observations which reveal
dramatic changes in the plasma's property while Galileo flying through the narrow wake of Io
($\sim 6\,R_I$) [Frank et al. 1996;  Kivelson et al. 1996; Gurnett et al. 1996]. The recent
high-quality images of [S {\small II}] 6731 \AA\ emissions from $S^+$, on the other hand, show
this plasma feature in unprecedented detail [Schneider and Trauger 1995].

Because we are mostly interested in its time-averaged properties, the ribbon can then be
approximated as in a steady state. We will assume that the narrow ribbon is surrounded by a
background plasma and the plasma and neutral gas distributions inside the ribbon only depend on
their azimuthal coordinate. By neglecting the variations of the gas and plasma properties in the
radial and vertical directions inside the ribbon, we reduce the study to an one dimensional
problem. As the first step of such a investigation, these approximations are good enough to
address the relevant energy transportation processes. However the detailed gas and plasma
distributions will control the corresponding diffusion and ionization processes, which are crucial
for a self-consistent hydrodynamical model. In the following discussion, we will parameterize
these processes in accord with these approximations.

We will develop the model in Io's rest frame. Then the mass conservation law is given by:
\begin{equation}
{\d\rho \over \d t }= {\partial\rho\over\partial t}+\nabla (\rho\mathbf{v})\,,
\end{equation}
where $\rho$ is the mass density of the ions inside the ribbon and $\mathbf{v} = v\mathbf{e_\phi}$
is the plasma's velocity. The neutral gas is roughly in a Keplerian orbit. Thus its velocity
in Io's rest frame is zero. Then we have
\begin{equation}
(\rho v)^\prime = {\rho_n\over \tau_n} - {\rho-\rho_b \over \tau_i}\,,
\label{mass-con}
\end{equation}
where $\rho_n$ is the neutral gas mass density, which is zero outside the ribbon, $\rho_b$ is the
background ion density, $\tau_n$ is the mean lifetime of the neutral gas, which characterizes the
ionization processes and $\tau_i$ is the diffusion time scale of the ions. A prime here denotes
derivative with respective to the azimuthal coordinate. In obtaining equation (\ref{mass-con}), we
have assumed that the background ions only have the effect of suppressing the diffusive loss of
ions from the ribbon.

From the momentum conservation equation
\begin{equation}
{\d \rho\mathbf{v}\over \d t} = {\partial \rho\mathbf{v}\over\partial t}+\nabla (\rho\mathbf{vv})\,,
\end{equation}
we have
\begin{equation}
jB-{\rho-\rho_b \over \tau_i}v = v(\rho v)^\prime +\rho v v^\prime\,,
\label{mom2}
\end{equation}
where $j$ is the current density and $B$ is the strength of the magnetic field. The first term on
the LHS of equation (\ref{mom2}) comes from the current acceleration by the magnetic field. The
second term is associated with the diffusive loss of ions from the ribbon. Combining this equation
with equation (\ref{mass-con}), we get
\begin{equation}
jB = {\rho_n\over \tau_n} v + \rho v v^\prime\,,
\label{mom-con}
\end{equation}
which says that the current acceleration provides the momentum to speed up the ion newly produced
via neutral gas ionization, which corresponds to the first term on the RHS, and the momentum to
accelerate the plasma flow as indicated by the second term on the RHS.

The energy conservation law is given by
\begin{equation}
{\d \rho \epsilon\over \d t} = {\partial  \rho \epsilon \over  \partial t} +
\nabla(\rho \mathbf{v}\epsilon)\,,
\end{equation}
where $\epsilon=v^2/2+\epsilon_{th}$ is the plasma energy density per unit mass and
$\epsilon_{th}$ is the plasma's thermal energy density per unit mass. From this, we have
\begin{equation}
jE-\Lambda-{\rho-\rho_b\over\tau_i}\left(v^2/2+\epsilon_{th}\right)
= (\rho v)^\prime(v^2/2+\epsilon_{th})+\rho v^2 v^\prime+\rho v \epsilon_{th}^\prime\,,
\label{eng-con}
\end{equation}
where $E = vB$ is the electric field, $\Lambda$ is the cooling rate due to line emission.
Multiplying equation (\ref{mom-con}) with $v$, we have
\begin{equation}
jE = {\rho_n\over \tau_n} v^2 + \rho v^2 v^\prime\,.
\label{cureng}
\end{equation}
Combining this equation with equation (\ref{eng-con}), one then gets
\begin{equation}
\Lambda = {\rho_n\over \tau_n}\left(v^2/2-\epsilon_{th}\right)-\rho v \epsilon_{th}^\prime\,.
\end{equation}
This equation shows the energy balance between the cooling due to line emission and the heating
introduced by the injection of new plasma produced through neutral cloud ionization, corresponding
to the first term on the RHS, and energy advection which corresponds to the second term on the
RHS.

To calculate the power carried by Birkeland current, which includes the power dissipated in the
Jovian ionosphere and that consumed in the electron accelerating double layers [Mauk et al. 2002],
we note that the e.m.f. per unit length produced by the plasma inside the ribbon equals $B v_0$,
where $v_0=57\,$km~s$^{-1}$ is the rigid corotation velocity at Io's orbit. Multiplying equation
(\ref{mom-con}) with $(v_0-v)$, we have
\begin{equation}
jE_J = {\rho_n\over \tau_n}  v (v_0-v) + \rho v v^\prime (v_0-v)\,,
\end{equation}
where $E_J=B(v_0-v)$ is the electric field in Jupiter's co-rotation frame. Then the LHS gives the
power of the Birkeland current. The first term on the RHS comes from the pickup current and the
second term is associated with the acceleration current.

Equation (\ref{mom-con}) can be written as
\begin{equation}
J = {\eta_n\over \tau_n} v + \eta v v^\prime\,,
\label{mom-con1}
\end{equation}
where $J = \int \d z\ j$ is the height-integrated current density, $\eta_n = \int \d z(\rho_n/B)$
and $\eta = \int \d z (\rho/B)$ are the neutral gas content and plasma content per unit magnetic
flux inside the ribbon respectively. This current closes in the Jovian ionosphere. Then the
current continuity equation can be written as [Hill 2002]
\begin{equation}
J = 2 J_\theta L^{-3/2}\,,
\label{continuity}
\end{equation}
where
\begin{equation}
J_\theta = \Sigma E_\theta
\label{ohm}
\end{equation}
is the height-integrated Peterson current, $\Sigma$ is the height-integrated conductivity
of Jovian ionosphere [Strobel et al. 1983], $E_\theta$ is the strength of the electric field in
Jovian ionosphere and $L= 1/\sin^2\theta$. If we neglect the potential drop across the double
layers in the Birkeland circuit, we have
\begin{equation}
E_\theta= -E_J (dL/d\theta) = 2E_J L^{3/2}\,.
\label{potential}
\end{equation}
Then from equations (\ref{mom-con1}), (\ref{continuity}), (\ref{ohm}), (\ref{potential}) and
(\ref{mass-con}), we have the basic equations for the ribbon:
\begin{eqnarray}
4\Sigma B (v_0-v) &=& {\eta_n\over \tau_n} v + \eta v v^\prime\,,
\label{mom-con2} \\
{\eta_n\over \tau_n} - {\eta-\eta_b\over \tau_i} &=& (\eta v)^\prime\,,
\label{mass-con1}
\end{eqnarray}
where $\eta_b = \int \d z(\rho_b/B)$ is the plasma content per unit magnetic flux of the
background within which the ribbon embedded. Then if we neglect the neutral gas injection and
plasma diffusion, i.e. let $\eta_n/ \tau_n = (\eta-\eta_b)/\tau_i=0$, we have the results
reported by Hill [2002]. Note in Hill's paper, he assumed that $\eta$ is a constant, which is
actually a very crude treatment of the plasma diffusion processes and the mass conservation law.
Our equations are accurate with the corresponding physical processes parameterized. On the other
hand, if we neglect the acceleration current, we have the results given by Hill and Pontius
[1998] and by Pontius and Hill [1982].

\section{Model Parameters and an Analytical Solution}
\label{para}

In equations (\ref{mom-con2}) and (\ref{mass-con1}), $\eta_n/\tau_n$ depends on the mechanism
through which the neutral gas is picked up from Io and gets ionized, and $\eta_b$ depends on how
the plasma diffuses away from the torus. Because both processes are not well understood, we will
leave $\eta_n/\tau_n$, $\eta_b$ as input functions. $B$ is measured by Kivelson et al. [1996] with
the magnetometer on Galileo: $B\sim 1400\,$nT. $\Sigma$ is given by aeronomy models [e.g., Strobel
and Atreya, 1983]. We will adopt a fiducial value of $0.1\,{\rm mho}$. For given $\eta_n$ and
$\eta_b$, $\tau_i$ can be fixed by the steady state condition (we will address this in detail in
the following discussion) when we solve the equations for $v$ and $\eta$. Then with the thermal
properties of the plasma determined with models of the corresponding microscopic processes
[Barbosa, Coroniti and Eviatar 1983], we can calculate the model predicated emissions from the
ribbon, which can be compared with observations directly. At the same time, the power of the
Birkeland current can be associated with observations on the Io's footprints on Jovian ionosphere
[Clarke et al 2002]. Strong correlation between emissions from the torus and that from Io's
footprints is expected.

To solve the equations numerically, it is convenient to put the equations in a dimensionless form.
We define
\begin{eqnarray}
v(\phi) &=& f(\phi) v_0\,, \\
{\eta_n\over \tau_n} &=& \lambda_0 \Lambda(\phi)\,, \\
\eta &=& \lambda_0\tau_0 g(\phi)\,, \\
\eta_b &=& \lambda_0\tau_0 g_b(\phi)\,,
\end{eqnarray}
where
\begin{equation}
\tau_0 = {L R_J\over v_0} = 7.3\times 10^3 {\rm sec}
\left({R_J\over 7.1\times 10^7{\rm m}}\right)
\left({L\over 5.9}\right)
\left({5.7 \times 10^4{\rm m\ sec}^{-1}\over v_0}\right)
\end{equation}
gives the characteristic dynamical time scale. $\lambda_0$ is defined as
\begin{equation}
\lambda_0 = {\dot{M}\over B 2\pi L R_J w}\,,
\end{equation}
where $\dot{M}$ is the total mass injection rate throughtout the ribbon, $L=5.9$ is orbital radius
of the torus [Broadfoot et al. 1979], $R_J= 7.1\times 10^7$ m is Jupiter's radius, $w$ is the
width of the ribbon. We will assume that the ribbon has a width of $6\,R_I$. Observations
[[Trauger 1984; Dessler and Sandel 1993; Schneider and Trauger 1995] indicate that the height of
the narrow bright feature in Io's plasma torus is $\sim R_J$, which is much larger than the width
of the ribbon. This is because the diffusion of ions is not isotropic in the plasma torus due to
the existence of the strong magnetic field, which is roughly perpendicular to the plasma's orbital
plane. Such an anisotropy will make the plasma spread out in the direction parallel to the
magnetic field line. And the deviation of the magnetic field line from the vertical direction also
enhances this effect when we have new ions produced via neutral gas ionization as is the case in
our model. Because the observed feature is associated with the ion distribution directly, we would
expect that the emission region be more extended in the vertical direction than in the radial
direction. However, the diffusion of neutral gas is almost isotropic. So we can assume that the
cross section of the ribbon is a square.

There are several independent estimates of the mass injection rate into the torus. The UVS
observation shows the total power of the UV emission from the torus is no less than $2.5\times
10^{12}$ W [Sandel et al., 1979], implying a mass loading rate $\dot{M} > 1.6\times 10^3$
kg sec$^{-1}$ [Siscoe and Chen, 1977]. Dessler [1980] show that $\dot{M} > 10^3$ kg sec$^{-1}$
based on the energy supply to the Jovian magnetosphere from Jupiter's rotation. The plasma
production via ionization of the neutral gas from Io in the magnetosphere will induce a slip
of the plasma relative to corotation as the plasma flows outwards and angular momentum is
transported from Jupiter out to the plasma flow [Hill 1979]. Voyager 1 plasma observations [Bridge
et al. 1979] indicate: $\dot{M}\sim 1.7\times 10^3 (\Sigma/0.1 {\rm mho})$ kg sec$^{-1}$ [Hill,
1980]. So $2.0\times 10^3$ kg sec$^{-1}$ is a reasonable fiducial mass injection rate.

Then we have
\begin{equation}
\lambda_0 = 5.0\times 10^{-8}
\left({\dot{M}\over 2\times 10^3{\rm kg\ s^{-1}}}\right)
\left({1400 {\rm nT}\over B}\right)
\left({5.9\over L}\right)
\left({7.1\times 10^7 {\rm m}\over R_J}\right)
\left({1.1\times 10^7 {\rm m}\over w}\right)
{\rm kg\ (sTm^2)^{-1}}\,.
\end{equation}
Equations (\ref{mom-con2}) and (\ref{mass-con1}) then can be written as:
\begin{eqnarray}
\sigma (1-f) &=& \Lambda f+ g f {\d f\over \d\phi}\,,
\label{mom-con3}\\
\Lambda - \alpha (g-g_b) &=& {\d (gf)\over \d \phi}\,,
\label{mass-con2}
\end{eqnarray}
where $\alpha = \tau_0/\tau_i$ and
$$\sigma ={4\Sigma B\over \lambda_0} = 11.3
\left({\Sigma\over 0.1 {\rm mho}}\right)
\left({B\over 1400{\rm nT}}\right)
\left({5.0\times 10^{-8} {\rm kg\ (sTm^2)^{-1}}\over \lambda_0}\right)\,.
$$

A self-consistent steady state solution is also subject to the following constraints:
\begin{eqnarray}
\int \d \phi R_J L\ w {\eta_n\over \tau_n} = {\dot{M}\over B}
&\iff& \int\d\phi\ \Lambda = 2\pi\,,
\label{massload}\\
\int \d \phi R_J L\ w \left({\eta_n\over \tau_n}-{\eta-\eta_b\over \tau_i}\right) = 0
&\iff& \int \d\phi\ \left[\Lambda-\alpha (g-g_b)\right] =0\,.
\label{balance}
\end{eqnarray}
The first constraint sets the normalization condition for $\Lambda$. To get the second constraint,
we have assumed that the flux of the plasma inside the ribbon satisfies the periodic boundary
condition. However, other physical properties of the plasma don't need to be continuous across Io
because shocks will be produced there. For given neutral cloud distribution in the torus,
$\Lambda(\phi)$, satisfying condition (\ref{massload}), background ion distribution $g_b$ and
boundary conditions $f(0)$ and $g(0)$, it is obvious that not all the solutions of equations
(\ref{mom-con3}) and (\ref{mass-con2}) satisfy condition (\ref{balance}). Only certain value of
$\alpha$ (eign value) can make condition (\ref{balance}) satisfied. For arbitrary input functions
$\Lambda(\phi)$ and $g_b$, we must choose an $\alpha$ to get a solution, then we check on
condition (\ref{balance}) and adjust $\alpha$ until it is satisfied. The calculation is quite
complicated.

Fortunately, in the case that $\Lambda-\alpha (g-g_b)=0$ everywhere and $g_b$ is a constant, which
we call local equilibrium in a smooth background, we have the following solution:
\begin{eqnarray}
f &=& {\sigma-c_1\alpha\over \sigma-\alpha g_{b}}\left(1-c_2 {\rm e}^{-(\sigma-\alpha g_{b})
\phi/c_1}\right)\,,
\\
g &=& c_1/f \,, \\
\Lambda &=& \alpha (g-g_b)\,,
\end{eqnarray}
where $c_1=gf$, $c_2 = 1-(\sigma-\alpha g_b) f(0)/(\sigma - \alpha c_1)$ can be fixed with the
boundary conditions $f(0)$ and $g(0)$. Galileo observation shows that the averaged mass
flux in Io's wake is about $<\rho v>\sim 1.4\times10^{-11}$ kg sec$^{-1}$ m$^{-2}$, where we have
used a background speed of $5.3\times 10^4$ m sec$^{-1}$ [Bagenal 1997].
We also assumed a mean ion mass of 24 amu [Hill 2002]. From this we have
\begin{eqnarray}
c_1 &=& { H <\rho v> \over v_0\tau_0\lambda_0 B} \nonumber \\
&=& 5.2
\left({5.7\times10^4{\rm m\ s^{-1}}\over v_0}\right)
\left({7.3\times 10^3 {\rm s}\over \tau_0}\right)
\left({5.0\times 10^{-8} {\rm kg\ (sTm^2)^{-1}}\over \lambda_0}\right)
\nonumber \\
&&
\left({1400 {\rm nT}\over B}\right)
\left({H\over w}\right)
\left({w\over 6R_I}\right)
\left({R_I\over 1.8\times 10^6 {\rm m}}\right)
\left({<\rho v>\over 1.4\times 10^{-11} {\rm kg\ s^{-1} m^{-2}}}\right)\,.
\label{c1}
\end{eqnarray}

Galileo observations also reveal a sharp increase of electron number density and decrease in the
flow velocity in Io's wake. And the background mass flux is $<\rho v>_b\sim 8.0\times10^{-12}$ kg
sec$^{-1}$ m$^{-2}$. We will adopt a value of $3.0\times 10^4$ m sec$^{-1}$ as a characteristic
averaged initial velocity of the flow inside the ribbon. Then the initial mean mass density of the
plasma is $4.8\times 10^{-16}$ km m$^{-3}$, corresponding to an ion number density of $12000$
cm$^{-3}$. Then we have \begin{eqnarray}
g_b &=& 3.2
\left({5.3\times10^4{\rm m\ s^{-1}}\over v_b}\right)
\left({7.3\times 10^3 {\rm s}\over \tau_0}\right)
\left({5.0\times 10^{-8} {\rm kg\ (sTm^2)^{-1}}\over \lambda_0}\right)
\nonumber \\
&&
\left({1400 {\rm nT}\over B}\right)
\left({H\over w}\right)
\left({w\over 6R_I}\right)
\left({R_I\over 1.8\times 10^6 {\rm m}}\right)
\left({<\rho v>_b\over 8.0\times 10^{-12} {\rm kg\ s^{-1} m^{-2}}}\right)\,,
\label{gb} \\
c_2 &=& 1-0.53\left[{v_i\over v_o}\right]\left({11.3[\sigma/11.3]-3.2\alpha[g_b/3.2]\over
11.3[\sigma/11.3]-5.2\alpha[c_1/5.2]}\right)\,.
\label{c2}
\end{eqnarray}

From the normalization condition (\ref{massload}) we have:
\begin{equation}
{\alpha c_1^2\over \sigma-c_1 \alpha}\left[{2\pi(\sigma-\alpha g_b)\over c_1} + \ln{(1-c_2{\rm
e}^{-2\pi(\sigma-\alpha g_b)/c_1})}-\ln {(1-c_2)}\right] = 2\pi(1+g_b\alpha)\,.
\label{alpha}
\end{equation}
Solving this equation for $\alpha$, we get an analytical solution for the structure of the ribbon.

\section{Model Fit to Observations}
\label{fit}

The power per unit magnetic flux carried by the Birkeland current is given by
\begin{equation}
P_B = \int \d z {E_J j\over B} = \lambda_0v_0^2(1-f)f[\Lambda+g(\d f/\d\phi)]\,.
\end{equation}
The corresponding dissipation inside the torus is given by
\begin{eqnarray}
P_t &=& \int \d z \left({\rho_n\over 2\tau_n B}v^2+{\rho-\rho_b\over 2\tau_i B}(v_b-v)^2\right)
\nonumber \\
&=& 0.5 \lambda_0v_0^2[\Lambda f^2 + \alpha (g-g_b)(v_b/v_0-f)^2]\,.
\end{eqnarray}
These powers can be compared with observations with appropriate models which can predict the
observed emission from the Jupiter-Io system. Fitting observations of the FUV Io tail on Jupiter
[G\'{e}rard et al. 2002], we find
\begin{eqnarray}
\Sigma &=& 0.15 {\rm mho}\,,\\
\sigma&=& 17\,.
\end{eqnarray}
Then solving equations (\ref{c2}) and (\ref{alpha}), one has
\begin{eqnarray}
\alpha &=& 0.41   \,,\\
c_2 &=& 0.44  \,.
\end{eqnarray}

\begin{figure}[thb]
{\begin{turn}{0}
\epsscale{0.8}
\centerline{\plotone{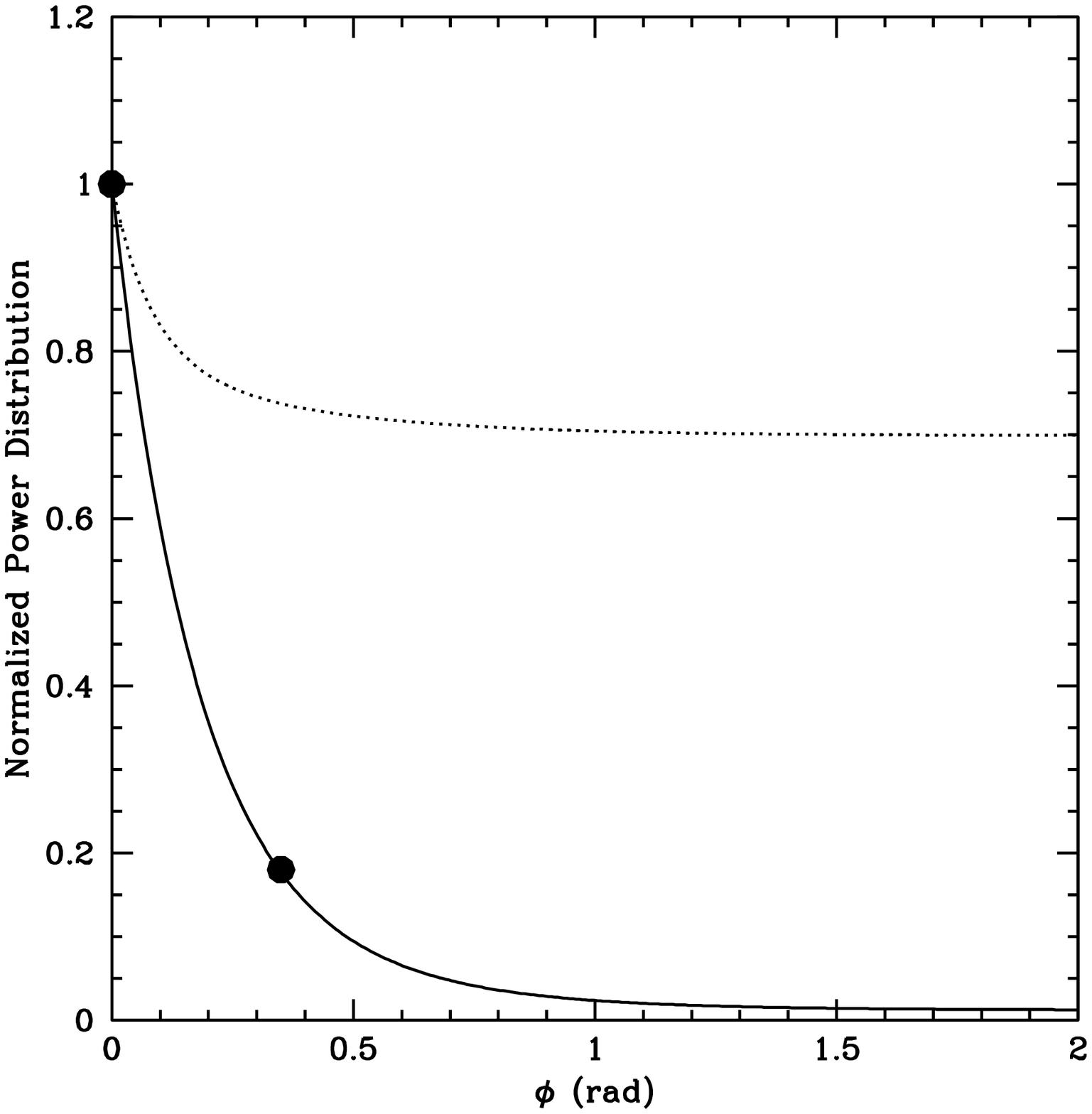}}
\end{turn}}
\caption
{
Normalized power distributions in the azimuthal direction. The solid line gives the power
distribution associated with the Birkeland current ($P_B$), which should be compared with
observations of the downstream emission of Io's footprint on Jupiter. The dotted line is for the
power dissipation in the torus ($P_t$). Both curves are normalized to their initial values which are
$3.9\times 10^{12}\,$W~rad$^{-1}$, and $6.1\times 10^{11}\,$W~rad$^{-1}$ respectively. The dots
correspond to the data inferred from observations of the FUV Io tail on Jupiter [G\'{e}rard et
al. 2002].
}
\label{fig1.ps}
\end{figure}

\begin{figure}[thb]
{\begin{turn}{0}
\epsscale{0.8}
\centerline{\plotone{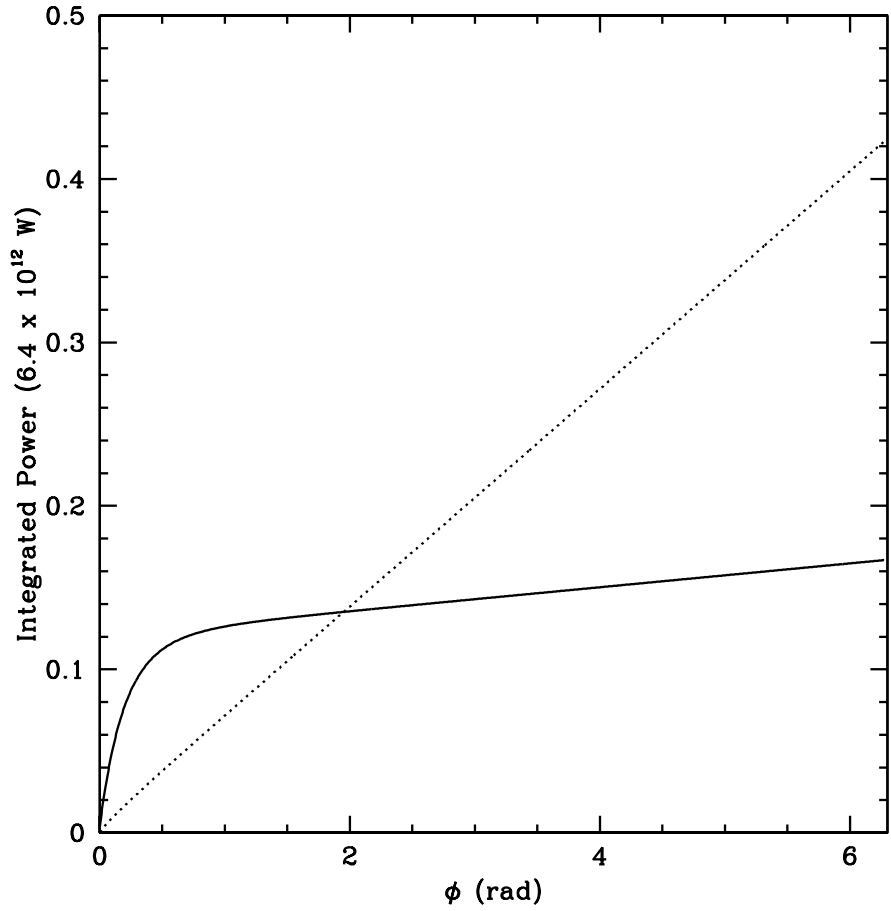}}
\end{turn}}
\caption{Integrated power distributions. The solid line is for the power associated with the
Birkeland current. The dotted line corresponds to the power dissipation in the torus.}
\label{fig2.ps}
\end{figure}

So we have the structure of the ribbon:
\begin{eqnarray}
f &=& 0.95-0.42 {\rm e}^{-3.0\phi} \,,\\
g &=& 5.2/(0.95-0.42 {\rm e}^{-3.0\phi}) \,,\\
\Lambda &=& 2.1/(0.95-0.42 {\rm e}^{-3.0\phi})-1.3\,,\\
P_B &=& 160\ (0.05+0.42{\rm e}^{-3.0\phi})\ (0.87+7.1{\rm e}^{-3.0\phi})
\nonumber \\
&&
\left({\lambda_0\over5.0\times 10^{-8}}\right)
\left({v_0\over 5.7\times 10^4{\rm m\ s^{-1}}}\right)^2 {\rm W\ m^{-2} T^{-1}}\,, \\
P_t
&=& 81\ \left({2.1\over 0.95-0.42 {\rm e}^{-3.0\phi}}-1.3\right)
\ \left(0.90-0.80{\rm e}^{-3.0\phi}+0.35 {\rm e}^{-6.0\phi}\right)
\nonumber \\
&&
\left({\lambda_0\over5.0\times 10^{-8}}\right)
\left({v_0\over 5.7\times 10^4{\rm m\ s^{-1}}}\right)^2 {\rm W\ m^{-2} T^{-1}}\,.
\end{eqnarray}
In figure \ref{fig1.ps} we give the normalized power distributions in the azimuthal direction.
The model gives a reasonable fit to observations of the FUV Io tail on Jupiter.
Figure \ref{fig2.ps} shows the integrated power distributions. The integrations start at Io.
From this we see that most of the power associated with the Birkeland current is localized around
Io. However, the power dissipation in the torus is quite uniform. The model then fails to explain
the Io phase effect which says that about $20\%$ of the EUV power from the Io's plasma torus is
correlated with Io [Sandel \& Broadfoot 1982b]. This is not a surprise given our crude treatment
of the boundary conditions near Io. Galileo {\it in situ} observations show that the power
supplied to the torus near Io accounts for $15\%$ to $28\%$ of the total EUV power from the torus
[Bagenal 1997], suggesting that Io phase effect is associated with the processes near Io.
Investigation of these processes is beyond the scope of this paper. A global self-consistent
picture of the Jupiter-Io system should incorporate these processes with the model we developed
here for the ribbon.

We also note that the total power dissipated in the torus is $2.7\times 10^{12}\,$W, which is
consistent with the EUV observations of emission from the torus. The power associated with the
Birkeland current is $1.1\times 10^{12}\,$W which, when combined with the observed FUV power from
Io's footprint, implies a FUV radiation efficiency of $\sim 10\%$. This total power supply to
Jovian ionosphere via Birkeland current sets a very strict constraint on the excitation mechanism
of the FUV Io tail. When combining with an auroral atmosphere model, one can infer an model
predicted $H_3^+$ brightness at Io's footprint, which can be compared with the
corresponding observations. The total power supplied to the torus via ionization of the neutral
cloud in the ribbon is given by
\begin{equation}
P= {1\over 2} \dot{M} v_b^2 = 3.2\times 10^{12}\,{\rm W}
\left({\dot{M}\over 2\times10^3{\rm kg\,s}^{-1}}\right)
\left({v_b\over 5.3\times 10^4{\rm m\,s}^{-1}}\right)^2\,,
\end{equation}
which is smaller than the sum of the power dissipated in the torus and that associated with the
Birkeland current. The power excess comes from the mass loading near Io, which breaks the
periodic conditions for the ribbon. So there is a net energy flux from the ends of the ribbon,
where Io locates.

\begin{figure}[thb]
{\begin{turn}{0}
\epsscale{0.8}
\centerline{\plotone{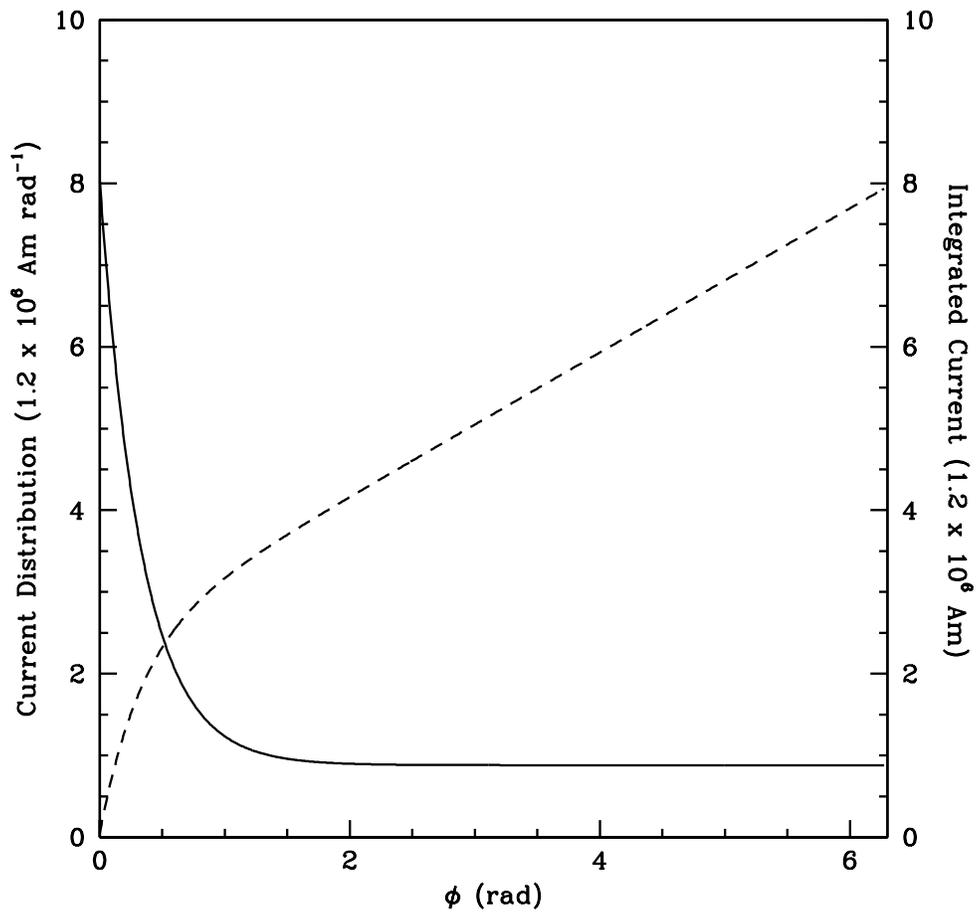}}
\end{turn}}
\caption{Current distribution (solid line, whose scale is on the left-hand side) and
integrated current (dashed line, which scales as the right-hand side).}
\label{fig3.ps}
\end{figure}

In figure \ref{fig3.ps}, we show the current distribution $J$ in the azimuthal direction, where
\begin{equation}
J = 1.2\times10^{6}(0.87+7.1{\rm e}^{-3.0\phi})\,{\rm Am\,rad}^{-1}\,,
\end{equation}
and the integrated current. The current near Io is $\sim 4\times 10^6\,$Ampere, which is
consistent with Voyager 1 {\it in situ} observations [Belcher 1987]. The total current
across the ribbon is $9.6\times10^6\,$Ampere. Because the newly produced ion via ionization of the
neutral cloud gains momentum through the current, we can estimate the total current across ribbon
from the mass injection rate in the torus:
\begin{equation}
I = {\dot{M} v_b\over w B} = 7.0\times 10^6{\rm Am}
\left({\dot{M}\over 2\times 10^3{\rm kg\,s}^{-1}}\right)
\left({v_b\over 5.3\times 10^4{\rm m\,s}^{-1}}\right)
\left({1400 {\rm nT}\over B}\right)
\left({1.1\times 10^7 {\rm m}\over w}\right)\,.
\end{equation}
The fact that this current is smaller than the value from our numerical calculation also
suggests local excess of mass injection near Io, which requires extra current to accelerate the
ions in the ribbon to the velocity of the background plasma.

\section{Conclusions}
\label{con}

In this paper, we develop a self-consistent model for the bright ribbon in Io's plasma torus. Most
of the model parameters were determined theoretically or via observations. Thus the model is
well constrained. To compare with observations of the FUV Io tail on Jupiter, we gave the simplest
analytical solution of the equations, which control the structure of the ribbon. The model
predicts a neutral cloud ionization rate as:
\begin{equation}
{\rho_n\over \tau_n} = 6.4\times 10^{-27}\, {\rm kg\,cm^{-3}s^{-1}}
\left({2.1\over0.95-0.42 {\rm e}^{-3.0\phi}}-1.3\right)\,,
\end{equation}
which can be combined with a model detailing the ionization processes of the neutral cloud to
make predictions on observations in the optical band [Smyth and Shemansky 1983].

The model also requires an ion distribution of
\begin{equation}
n_i = {6400\over 0.95-0.42{\rm e}^{-3.0\phi}}\, {\rm cm^{-3}}\,,
\end{equation}
inside the ribbon. This number density is higher than the canonical value of $2000\,$cm$^{-3}$ by
more than a factor of 3. Although this high number density implies a relatively larger cooling
rate, the density is not high enough to explain the short cooling time reported by Volwerk et al.
[1997]. However, we noticed that the lack of correlation between the brightnesses of the two ansae
can also be explained as due to some plasma processes, such as plasma waves et al. These processes
can dump a local enhancement of energy density with the corresponding time scales. For example,
the Alfv\'{e}n velocity inside the ribbon is more than two times bigger than the co-rotation
velocity at Io. The corresponding time scale is more than two times smaller than the corresponding
dynamical time scale which is enough to explain the observed the lack of correlation. Further
investigation of its implication on the EUV emission processes is warranted [Barbosa, Coroniti,
and Eviatar 1983a].

{\bf Acknowledgments}

{}

%
%

\end{document}